\documentclass{PoS}

\title{NIKA2: a mm camera for cluster cosmology}

\ShortTitle{NIKA2: a mm camera for cluster cosmology}

\author{\speaker{J.F.~Mac\'{\i}as-P\'erez},
R.~Adam$^{1,2}$,
P.~Ade$^3$,
P.~Andr\'e$^4$,
M.~Arnaud$^4$,
H.~Aussel$^4$,
I.~Bartalucci$^4$,
A.~Beelen$^5$,
A.~Beno\^it$^6$,
A.~Bideaud$^3$,
O.~Bourrion$^1$,
M.~Calvo$^6$,
A.~Catalano$^1$,
B.~Comis$^1$,
M. De Petris$^7$,
F.-X.~D\'esert$^{8}$,
S.~Doyle$^3$,
E.~F.~C.~Driessen$^9$,
J.~Goupy$^6$,
C.~Kramer$^{10}$,
G.~Lagache$^{11}$,
S.~Leclercq$^9$,
J.~F.~Lestrade$^{12}$,
P.~Mauskopf$^{3,13}$,
F.~Mayet$^1$,
A.~Monfardini$^6$,
L.~Perotto$^1$,
E.~Pointecouteau$^{14}$,
G.~Pisano$^3$,
N.~Ponthieu$^{8}$,
G.~W.~Pratt$^4$,
V.~Rev\'eret$^4$,
A.~Ritacco$^{10}$,
C.~Romero$^9$,
H.~Roussel$^{15}$,
F.~Ruppin$^1$,
K.~Schuster$^9$,
A.~Sievers$^{10}$,
C.~Tucker$^3$,
R.~Zylka$^9$ \thanks{{\small We would like to thank the IRAM staff for their support during the campaigns. 
The NIKA dilution cryostat has been designed and built at the Institut N\'eel. 
In particular, we acknowledge the crucial contribution of the Cryogenics Group, and 
in particular Gregory Garde, Henri Rodenas, Jean Paul Leggeri, Philippe Camus. 
This work has been partially funded by the Foundation Nanoscience Grenoble, the LabEx FOCUS ANR-11-LABX-0013 and 
the ANR under the contracts "MKIDS", "NIKA" and ANR-15-CE31-0017. 
This work has benefited from the support of the European Research Council Advanced Grant ORISTARS 
under the European Union's Seventh Framework Programme (Grant Agreement no. 291294).
We acknowledge fundings from the ENIGMASS French LabEx (R. A. and F. R.), 
the CNES post-doctoral fellowship program (R. A.),  the CNES doctoral fellowship program (A. R.) and 
the FOCUS French LabEx doctoral fellowship program (A. R.).}}\\
        E-mail: \email{macias@lpsc.in2p3.fr}}

\author{\\  $^{1}$ Laboratoire de Physique Subatomique et de Cosmologie, Universit\'e Grenoble Alpes, CNRS/IN2P3, 53, avenue des Martyrs, Grenoble, France \\
 $^{2}$Laboratoire Lagrange, Universit\'e C\^ote d'Azur, Observatoire de la C\^ote d'Azur, CNRS, Blvd de l'Observatoire, CS 34229, 06304 Nice cedex 4, France\\
 $^{3}$Astronomy Instrumentation Group, University of Cardiff, UK \\
$^{4}$Laboratoire AIM, CEA/IRFU, CNRS/INSU, Universit\'e Paris Diderot, CEA-Saclay, 91191 Gif-Sur-Yvette, France\\
 $^{5}$Institut d'Astrophysique Spatiale (IAS), CNRS and Universit\'e Paris Sud, Orsay, France\\
 $^{6}$Institut N\'eel, CNRS and Universit\'e Grenoble Alpes, France\\
 $^{7}$Dipartimento di Fisica, Sapienza Universit\`a di Roma, Piazzale Aldo Moro 5, I-00185 Roma, Italy\\
 $^{8}$Institut de Plan\'etologie et d'Astrophysique de Grenoble, Univ. Grenoble Alpes, CNRS, IPAG, 38000 Grenoble, France\\ 
 $^{9}$Institut de RadioAstronomie Millim\'etrique (IRAM), Grenoble, France\\
 $^{10}$Institut de RadioAstronomie Millim\'etrique (IRAM), Granada, Spain\\
 $^{11}$Aix Marseille Universit\'e, CNRS, LAM (Laboratoire d'Astrophysique de Marseille) UMR 7326, 13388, Marseille, France\\ 
 $^{12}$LERMA, CNRS, Observatoire de Paris, 61 avenue de l'observatoire, Paris, France\\
 $^{13}$School of Earth and Space Exploration and Department of Physics, Arizona State University, Tempe, AZ 85287\\
 $^{14}$Universit\'e de Toulouse, UPS-OMP, Institut de Recherche en Astrophysique et Plan\'etologie (IRAP), Toulouse, France\\
 $^{15}$Institut d'Astrophysique de Paris, Sorbonne Universit\'es, UPMC Univ. Paris 06, CNRS UMR 7095, 75014 Paris, France
 }

\abstract{Galaxy clusters constitute a major cosmological probe. However, Planck 2015 results have shown a weak tension between CMB-derived and cluster-derived cosmological parameters. This tension might be due to poor knowledge of the cluster mass and observable relationship.
 As for now, arcmin resolution Sunyaev-Zeldovich (SZ) observations ({\it e.g.} 
SPT, ACT and Planck) only allowed detailed studies of the intra cluster medium for low redshift clusters ($z<0.2$). For high redshift clusters ( $z>0.5$) high resolution and high sensitivity SZ observations are needed.
With both a wide  field of view (6.5 arcmin) and a high angular resolution (17.7 and 11.2 arcsec at 150 and 260 GHz), 
the NIKA2 camera installed at the IRAM 30-m telescope (Pico Veleta, Spain) is particularly well adapted for these observations. The
NIKA2 SZ observation program will map a large sample of clusters (50) at redshifts
between 0.5 and 0.9. As a pilot study for NIKA2, several clusters of galaxies have been observed with the pathfinder, NIKA, at the
IRAM 30-m telescope to cover the various configurations and observation
conditions expected for NIKA2.}

\FullConference{EPS-HEP 2017, European Physical Society conference on High Energy Physics\\
		5-12 July 2017\\
		Venice, Italy}

\begin{document}

\section{The NIKA2 camera at the IRAM 30-m telescope}
NIKA2 is a millimetre camera~\cite{Calvo16,NIKA2-Adam}, made of Kinetic Inductance Detectors (KID) and operated at 150 mK, which 
has been installed in September 2015 at the IRAM 30-m telescope.
NIKA2 observes the sky at 150 and 260 GHz with a wide field of view (FOV), 6.5 arcmin (2896 detectors),  high-angular resolution 
(17.7 and 11.2 arcsec, respectively),  and state-of-the-art sensitivities.
Prior to NIKA2, the NIKA camera~\cite{Monfardini11,Bourrion12,Calvo13,Monfardini13,Catalano:2014nml}  was operated at the IRAM 30-m telescope from 2012 to 2015 with a smaller field of view  (1.8 arcmin) due to the reduced number of KIDs (356). 

The NIKA2 camera, and equivalent the NIKA camera, is  well suited for high-resolution Sunyaev-Zeldovich (SZ) observations of cluster of galaxies. First, the 150~GHz channel is optimal for SZ measurements (large negative SZ spectral distortion) and the 260~GHz one (weak positive SZ spectral distortion) can be used to mitigate contamination by dusty galaxies. Second, the 30-m telescope high resolution and the camera large FOV allows us to map intermediate and high redshift clusters from the core up to their characteristic radius 
$R_{500}$. Third, the NIKA2 sensitivity in terms of Compton parameter is expected to be of the order of 
$10^{-4}$ per hour and per beam, allowing us to obtain reliable SZ mapping at high signal-to-noise ratio in a few hours per cluster.

\section{Why high-resolution thermal SZ imaging of clusters?}
The thermal SZ (tSZ) effect~\cite{Sunyaev:1972,Sunyaev:1980vz,Birkinshaw:1998qp} is due to the inverse Compton scattering of CMB photons on hot electrons of the intra-cluster medium. The tSZ effect induces  spectral distortions in the CMB spectrum, such that it is negative (positive) below (above) 217~GHz. This characteristic spectral feature can be used both to blindly detect cluster of galaxies and to map them in details. Indeed, the thermal SZ Compton parameter, $y$, is a direct measurement of the cluster electronic pressure integrated along the line-of-sight.
Thermal SZ measurements of clusters are usually combined with X-ray  observations, for which the surface brightness is related to the electronic density square,  
in order to probe the clusters physical properties. 

Clusters of galaxies, which are the largest gravitational bound objects in the universe, constitute cosmological observables of choice. Indeed, their distribution in mass and redhift, as well as their spatial distribution, are sensitive to the cosmological parameters describing the main properties
of the universe~\cite{Ade:2013lmv,PlancktSZspec,Ade:2015fva,Planckymap}.
In the last decade arcmin resolution experiments like the Planck Satellite~\cite{Ade:2015gva,Ade:2013skr}, the Atacama Cosmology Telescope 
(ACT)~\cite{Hasselfield:2013wf} and the South Pole Telescope (SPT)~\cite{Bleem:2014iim} have provided large catalogue of blindly tSZ-detected cluster of galaxies and have opened the path towards SZ cluster cosmology. 
In particular, recent Planck results have shown that there is weak tension~\cite{Ade:2015fva,Planckymap} between CMB and cluster estimates of the matter density $\Omega_M$
and the amplitude of the matter density fluctuations, $\sigma_8$. 
This discrepancy may be due to an incorrect estimation of the total mass of the cluster, via the
mass-observable relation. Among others this relation may depend on the redshift, the internal
structure and/or the dynamical state of the considered clusters. 
These effects are expected to be particularly important for high redshift clusters (z > 0.4) that can not be studied in details with arcmin resolution observations.
Therefore, high-resolution tSZ imaging of high-redshift cluster is needed to study their intra-cluster properties, such as dynamical states (mergers) and morphology (departure from sphericity) and their implications in the mass-tSZ scaling law. 

%

\section{SZ observations with the NIKA camera}

The NIKA camera was used as a pathfinder for NIKA2 in order to demonstrate the possibility to use large arrays of KIDs for general purpose millimeter astronomy~\cite{Catalano:2014nml,Ritacco:2016due,Bracco}. 
Furthermore, an extensive program was carried out for SZ science that permitted the high signal-to-noise mapping of five high redshift clusters of galaxies~\cite{Adam:2013ufa,Adam:2014wxa,Adam:2015bba,Adam:2016abn,Ruppin:2016rnt,Adam:2017mlj}. The NIKA SZ data 
combined with X-ray data allowed us to study the radial distribution of the thermodynamical properties of their intra-cluster medium, {\it i.e.}  pressure, density, temperature, 
and entropy radial profiles. The various SZ observations with NIKA were:

\noindent {\small $\bullet$} RX J1347.5-1145, a well-known
and strong tSZ source,   allowed us 
to perform the first SZ cartography ever achieved with a KID-based camera~\cite{Adam:2013ufa}. In particular, we confirm that this cluster 
is an ongoing merger.
 
\noindent {\small $\bullet$} CL J1226.9+3332, a massive high-redshift cluster, has been 
shown to present cluster parameters, mass and integrated Compton parameter,  consistent with the Planck best-fit scaling relation~\cite{Ade:2013lmv} obtained with a sample of nearby clusters. Although no conclusion can be drawn from a single high-redshift cluster, 
it highlights  the interest of the NIKA2 SZ large program that is dedicated to the observation of a cluster sample with redshift up to $0.9$.

\noindent {\small $\bullet$}   MACS J1423.9+2404, which has been used to investigate the impact of the contamination by radio and/or dusty point sources in the cluster field~\cite{Adam:2015bba}.

\noindent {\small $\bullet$}  MACS J0717.5+3745, for which we reported 
the first mapping of the kinetic SZ effect in a cluster~\cite{Adam:2016abn} 
as well as the first mapping of the hot gas temperature using X-ray and SZ imaging~\cite{Adam:2017mlj}, 
providing an independent cross-calibration of X-ray spectroscopic measurements.

\noindent {\small $\bullet$} PSZ1 G045.85+57.71, which has been used to demonstrate the possibility to use NIKA2 for a high-resolution follow-up of tSZ Planck-discovered clusters and to measure their physical properties~\cite{Ruppin:2016rnt}.


\section{The NIKA2 SZ large program}
The NIKA2 SZ large program, which corresponds to 300 hours of NIKA2 guaranteed time,
aims at observing a representative sample of tSZ detected high redshift clusters.
The NIKA2 SZ cluster sample consists of 50 clusters with redshift ranging between 0.4 and 0.9, which were selected from the Planck~\cite{Ade:2015gva,Ade:2013skr} and ACT tSZ catalogues~\cite{Hasselfield:2013wf}. The sample spans about one order of magnitude in mass assuming standard cluster tSZ-mass scaling laws \cite{Ade:2013lmv} and give an homogeneous coverage of the mass range.
The excellent  sensitivity of NIKA2 \cite{NIKA2-Adam} will allow us to obtain reliable tSZ  
mapping of these clusters of galaxies in only few hours (1 to 10 hours). 
The NIKA2 tSZ data combined with ancillary data (X-ray, optical and radio) will allow the study of the 
thermodynamic properties of the ICM, and will lead to significant improvements on the cosmological use of clusters of galaxies. The main goals of the program are to measure:

\noindent {\small $\bullet$} the thermodynamic properties of the cluster (temperature, entropy and mass radial profiles), 

\noindent {\small $\bullet$} the redshift evolution of the scaling law and of the cluster pressure profiles up to high redshift, 

\noindent {\small $\bullet$}  the cluster morphology and the dynamical state at high redshift 
(departure from spherical symmetry, merging events, cooling processes).

\end{document}